\documentclass[%
 reprint,
superscriptaddress,
longbibliography,% make titles show in bib
showpacs,%preprintnumbers,
nofootinbib,
 amsmath,amssymb,aps,
prd,
]{revtex4-2}

\nocite{TitlesOn}
\usepackage{lipsum} % Package to generate dummy text throughout this template
\newcommand{\cmmnt}[1]{}
\usepackage{bm}
\usepackage{physics}
\usepackage{comment}
\usepackage{floatrow} % for caption beside figure
\pdfoutput=1
\usepackage{amsmath,amsfonts,amsthm}
\usepackage{esdiff}  
\usepackage{booktabs}  % For better-looking tables
\usepackage{url}  % To make clickable url's
\usepackage{hyperref}  % for footnotes?
\usepackage{relsize}
\usepackage{cleveref}  % Detects type of ref when you write \cref. 
	\crefname{equation}{equation}{equations}
	\crefname{figure}{figure}{figures}	
	\crefname{table}{table}{tables}
\usepackage[caption=false]{subfig}

\usepackage[normalem]{ulem}

\usepackage[usenames,dvipsnames,svgnames,table]{xcolor}

\usepackage{graphicx}

\usepackage{aurical}
\usepackage[T1]{fontenc}
\usepackage[sc]{mathpazo} % Use the Palatino font
\usepackage[T1]{fontenc} % Use 8-bit encoding that has 256 glyphs
\usepackage{microtype} % Slightly tweak font spacing for aesthetics

\usepackage{booktabs} % Horizontal rules in tables
\usepackage{float} % Required for tables and figures in the multi-column environment - they need to be placed in specific locations with the [H] (e.g. \begin{table}[H])
\usepackage{hyperref} % For hyperlinks in the PDF

\usepackage{lettrine} % The lettrine is the first enlarged letter at the beginning of the text
\usepackage{paralist} % Used for the compactitem environment which makes bullet points with less space between them

\definecolor{darkred2}{HTML}{880808}
\definecolor{crimson}{HTML}{DC143C}

\usepackage{titlesec} % Allows customization of titles
\renewcommand\thesection{\Roman{section}} % Roman numerals for the sections
\renewcommand\thesubsection{\Alph{subsection}} % Alphabet for subsections
\titleformat{\section}[block]{\large\scshape\centering\bfseries}{\thesection.}{1em}{} % Change the look of the section titles

\titleformat{\subsection}[block]{\scshape\centering}{\thesubsection.}{1em}{} % Change the look of the section titles

\DeclareCaptionFormat{myformat}{#1#2#3\hrulefill}
\usepackage{float}
\floatstyle{plaintop}
\restylefloat{table}

\begin{document}
\nocite{TitlesOn}

\title{I Murdered Conan O'Brien and Nobody Will Ever Know \\\textit{an exercise in inference sabotage}}

\author{Eve Armstrong\thanks{earmstrong@amnh.org}}
\email{earmstrong@amnh.org -- or perhaps I am framing her.}
\affiliation{Department of Physics, New York Institute of Technology, New York, NY 10023, USA}
\affiliation{Department of Astrophysics, American Museum of Natural History, New York, NY 10024, USA}
\affiliation{\url{http://www.amazon.com/author/evearmstrong}}

\date{April 1, 2023}

\begin{abstract}
I employ an optimization-based inference methodology together with an Ising model, in an intentionally ineffectual manner, to get away with murdering an obstreperous scientific collaborator.  The antics of this collaborator, hereafter "Conan O'Brien," were impeding the publication of an important manuscript.  With my tenure date looming, I found myself desperate.  Luckily, I study inference, a computational means to find a solution to a physical problem, based on available measurements (say, a dead body) and a dynamical model assumed to give rise to those measurements (a murderer).  If the measurements are insufficient and/or the model is incomplete, one obtains multiple "degenerate" solutions to the problem.  Degenerate solutions are all equally valid given the information available, and thus render meaningless the notion of one "correct" solution.  Typically in scientific research, degeneracy is undesirable.  Here I describe the opposite situation: a quest to \textit{create} degenerate solutions in which to cloak myself.  Or even better: to render measurements incompatible with a solution in which I am the murderer.  Moreover, I show how one may sabotage an inference procedure to commit an untraceable crime.  I sit here now, typing victoriously, a free woman.  Because you won't believe me anyway.  And even if you do, you'll never prove a thing.  
\end{abstract}

\maketitle

\section{Introduction} \label{sec:intro} 

Toward the end of November this past year, I found myself increasingly fretful.  Our research group was on the brink of publishing an important manuscript on neutrino~\cite{pauli1930liebe}  flavor transformation~\cite{duan2010collective} in core-collapse supernovae~\cite{baade1934super} -- the violent explosion of a supermassive star at the end of its lifetime; very exciting.  We had submitted the manuscript to a respected peer-reviewed journal, and they had requested only minor revisions.  Delightful news, Dear Reader.

But for a problem.  

One collaborator in our group had a history of making incessant and contradictory nit picks that would contort a straightforward operation into a Gordeon knot.  This manuscript was no exception.  The collaborator would rehash the interpretation of our results, ad nauseam, teasingly backpedaling on long-accepted pillars of physics, such as whether it really does hurt to fall out of a tree.  Regarding the writing, he insisted that we replace American English with British English (e.g. "flavor" to "flavour.")  We did, and then he made the reverse demand, not remembering his original.  On the seventh draft he declared that he no longer believed in semicolons. 
% the {\color{purple} attractive property of gravity, existence of matter, existence of something like paper?, whether objects move, 

In addition, this collaborator considered himself something of a comedian.  He would do impersonations of neutrinos.  He even had variations that distinguished among muon, tau, and electron flavors.  Beyond "science," he made cat noises at group meetings.  Sometimes after a student's presentation, instead of clapping he'd purr\footnote{We've dodged bullets from the Office of Sexual Misconduct Prevention after some students -- understandably -- misinterpreted the purring.}.  One time he rolled around under the table batting a feather for five minutes, while we were hosting a visiting scholar from Harvard.  We couldn't get anything done.  And you should have seen his antics at restaurants; never have I felt so sorry for waiters\footnote{And I've been to Boston on Saint Patrick's Day.}.  Hereafter, for the purpose of obscuring my crime, I shall refer to this collaborator with an unremarkable name that would fit just about anybody.  Let's go with Conan O'Brien~\cite{conan}. 

Back to November.  The year's end was looming and I still had zero publications to show for it, and I'm coming up for tenure\footnote{Tenure is the coveted employment status in academia that unburdens faculty of having to continue doing their job well in order to get paid for it.  For the interested reader: some~\cite{helfand} see a problem with this practice.}.  I tried to reason with Conan O'Brien; we all did.  But he didn't seem to possess the neural circuitry to grasp our concerns\footnote{He only displayed consternation that his neutrino impressions were not better received, resentfully noting that geniuses are rarely appreciated as such during their lifetimes.}.  I was approaching the end of my rope.  Conan O'Brien needed to vanish, and clearly he wouldn't do that on his own.  It was up to me\footnote{If you aren't on my side by now, I don't know what more to tell you.}.  My only reservation was getting caught.

Conveniently, I understand how murder investigations work: in our supernova studies, we apply optimization-based inference.  Inference identifies the measurements required to find a unique solution to a problem, within the context of a physical model that is assumed to have given rise to those measurements.  It is a cousin of machine learning and artificial intelligence.  Used across geophysics~\cite{kalnay2003atmospheric}, neurobiology~\cite{armstrong2020statistical}, and astrophysics~\cite{laber2023inference}, this methodology has a strong record of cross-disciplinary applications.  Could I add yet another?  Could I apply optimization to thwart a murder investigation?
 
What if I were to turn this quest for uniqueness inside out, such that I cannot be identified uniquely as the murderer?  That is, either: 1) allow that I might be the murderer, but create multiple other ("degenerate") murderers so as to eliminate uniqueness, or 2) rejigger either model or measurements so that -- while a unique murderer might exist -- it cannot be me?  For readers who find themselves in a similar predicament, in what follows I shall describe how to exploit this sound and well regarded methodology to achieve a nefarious -- and I daresay justified -- aim.  

\section{Landscape: measurements and model}

A criminal investigation considers two components: measurements and model.  Measurements constitute the evidence that a murder occurred: body\footnote{Do not confuse the body with the body's \textit{identity}.  For example, the body does not necessarily have a head.}, bodily properties including visible wounds, location, time of death, weapon, and articles possibly discarded by the murderer, like a mug of Dwight D. Eisenhower.  

Unfortunately, criminal investigators will assume that some underlying dynamical system gave rise to the measurements.  The system in question will be a network of interacting entities, or suspects, at least one of whom is the murderer.  The dynamics governing these entities is a model, whose "model space" contains the set of all possible explanations for the measurements.  The measurements live in a sub-region of this model space in which such measurements are possible.  That is, the measurements will tug investigators to the region of the model space most likely to contain the solution to the murder at hand.  

\begin{figure}[htb]
\includegraphics[width=\textwidth]{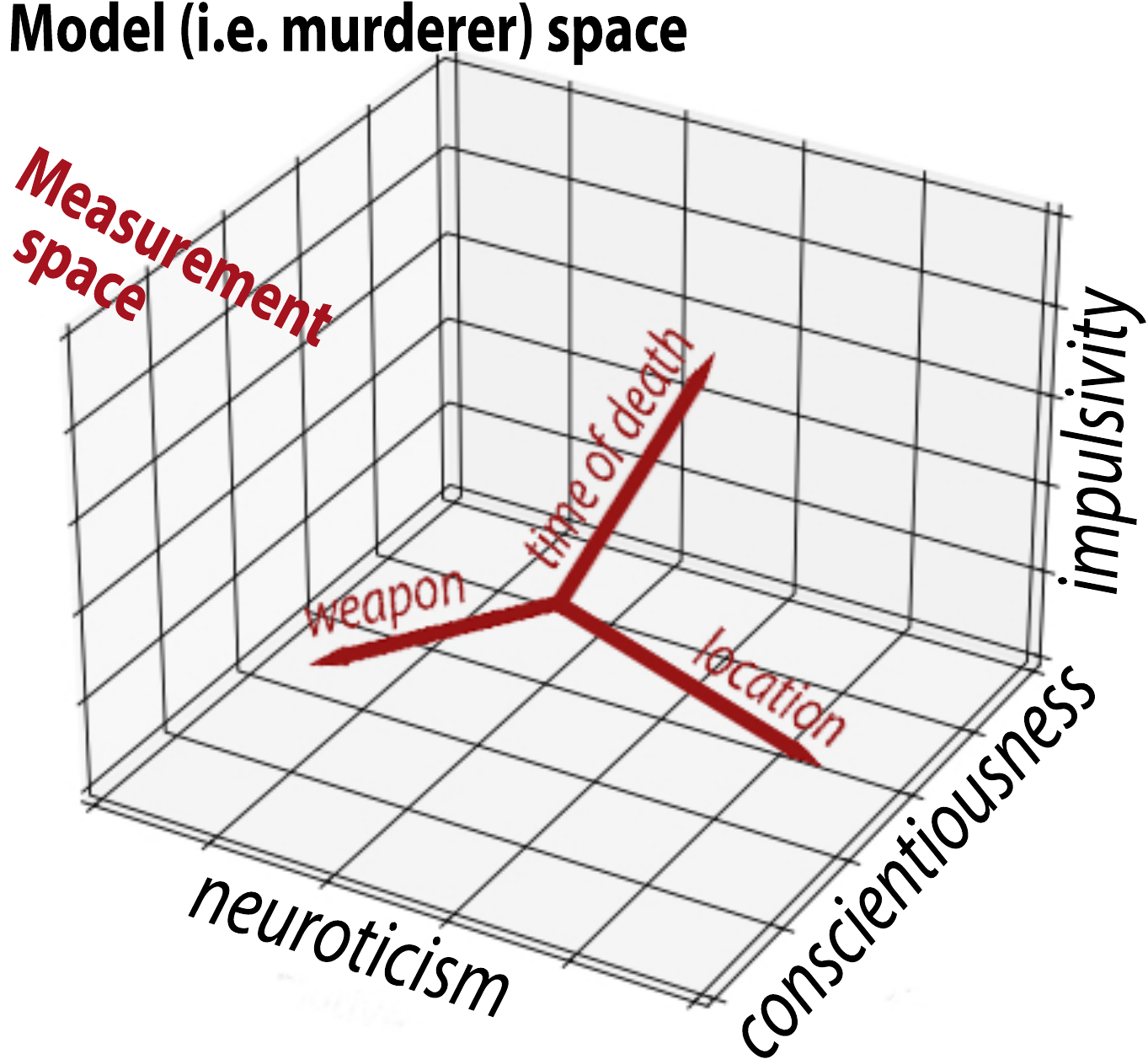}
\caption{\textbf{Three-dimensional slices of model space (black) and measurement subspace (crimson).}  Note: the two orthogonal sets of axes have been drawn arbitrarily with respect to each other; investigators will have to construct for themselves the proper translation between measurement and model dimensions.} 
\label{fig1}
\end{figure}

Fig.~\ref{fig1} depicts the model and measurement spaces.  For visualization, just three dimensions are shown for each.  For the model space (black), these are personality traits of the possible murderers: [neuroticism, impulsivity, conscientiousness].  The measurement subspace (crimson) is depicted in the dimensions of [weapon, location,

\noindent  time].  Note that a transfer function will be required to map from measurement to model dimensions (Sec.~\ref{sec:opt}).

Now, good news for prospective murderers who wish to remain anonymous: measurements may contain uncertainties (noise), and may be missing or inaccurate.  Thus they can be readily manipulated to tug investigators in multiple directions at once, or the wrong direction entirely (Sec.~\ref{sec:sabotage}).

\section{Model}

Let us consider a dynamical system $\bm{X}$ consisting of $N$ interacting entities $\bm{x}_i$, one of whom is Conan O'Brien ($\bm{x}_{\text{Conan}}$) and at least one of whom is the murderer ($\bm{x}_{murderer}$).  Generally, this system evolves in time $t$ according to some model $\bm{F}$, which we represent in $B$ ordinary differential equations\footnote{A more realistic model would also vary spatially.  E.g. I feel the most vexed by Conan O'Brien within close proximity to him.  Partial differential equations are more cumbersome, however, so let us take the chance that the police won't dig into that level of detail.}:
\begin{equation} \label{eq:ODE}
  \diff{x_b(t)}{t} = F_b(\bm{X}(t),\bm{p}(t)); \hspace{1em} b =1,2,\ldots,B,
\end{equation}

\begin{figure*}[htb]  
  \includegraphics[width=\textwidth]{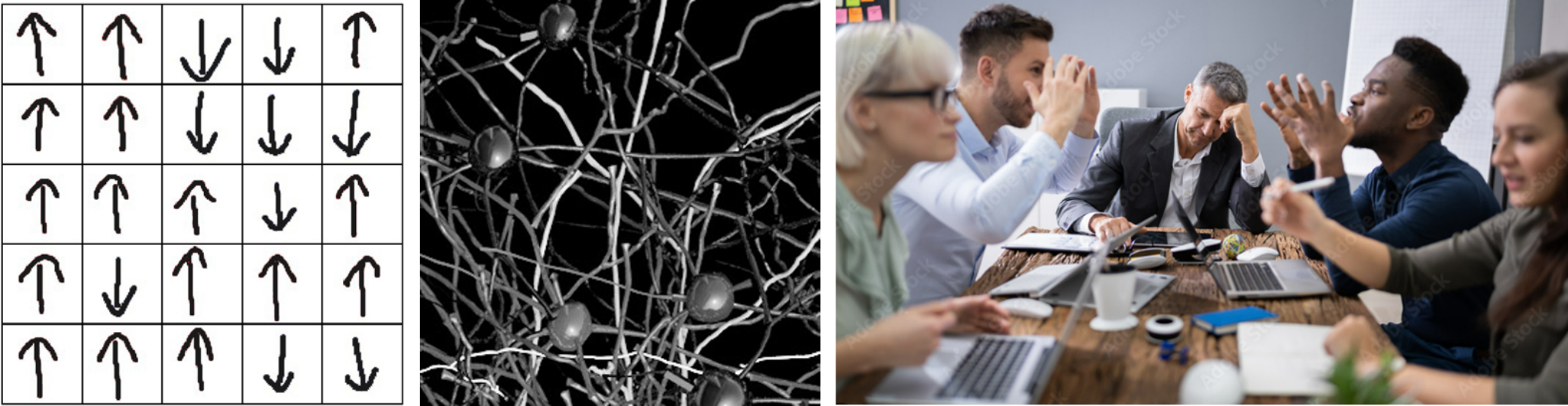}
  \caption{\textbf{Systems of coupled oscillators.}  \textit{Left}: Ising model of ferromagnetism~\cite{isingImage}, with spins $x_a$.  \textit{Middle}: Neuronal network~\cite{allenInst}, where individual neurons $\bm{x}_i$ are defined in terms of electrophysiological properties $x_{i,a}$.  \textit{Right}: Faculty meeting~\cite{facultyMeeting}, where potential murderers $\bm{x}_i$ are defined in terms of psychological and physiological traits $x_{i,a}$.} \label{fig2}
\end{figure*}

\noindent
where the $x_b$ are state variables that define the system $\bm{X}$, $F_b$ is the component of $\bm{F}$ that operates on state variable $x_b$, and $\bm{p}(t)$ are any unknown model parameters.  Note that these are Markov-chain~\cite{markov1906extension}, or memoryless\footnote{This might explain why so many murderers deny their crimes: maybe they forgot.}, dynamics, where the probability of an event occurring at time $t$ is determined entirely by the model state at the previous time $(t-1)$.

In its simplest form, a system of interacting entities is the Ising model of ferromagnetism~\cite{ising1925contribution}.  The Ising model is a lattice of magnetic dipole moments of atomic spins that can exist in one of two discrete states (Fig.~\ref{fig2}, left panel).  More complex formulations have since been written to describe neuronal networks~\cite{ermentrout2019recent} (Fig.~\ref{fig2}, middle), quantum mechanical oscillators~\cite{brown2011coupled}, cancer~\cite{feillet2014phase}, ant colonies~\cite{boi1999coupled}, and human romantic relationships~\cite{zee2020using}.  Within all these contexts, the respective populations are taken to be coupled oscillators.  Given its clear applicability across disciplines, let us extend the Ising model to a system of possible murderers (Fig.~\ref{fig2}, right).

With that aim, we shall write model $\bm{F}$ in two terms, where entities $\bm{x}_i$ of population $\bm{X}$ evolve as: 
\begin{equation} \label{eq:ising}
\diff{\bm{x}_{i}(t)}{t} = \underbrace{\bm{R}_i(\bm{x}_i)}_{\text{self}} + \underbrace{\sum_{j \neq i} \bm{S}_{ij}(\bm{x}_i,\bm{x}_j)}_{\text{interaction}}.
\end{equation}  
\noindent Here, the "self" term $\bm{R}_i(\bm{x}_i)$ describes the intrinsic properties of entity $\bm{x}_i$, and each "interaction" term $\bm{S}_{ij}(\bm{x}_i,\bm{x}_j)$ describes the dynamics that emerge when entity $\bm{x}_i$ encounters entity $\bm{x}_j$.  The $\bm{R}_i$ and $\bm{S}_{ij}$ terms contain the unknown parameters $\bm{p}$ of Eq.~\ref{eq:ODE}.  Those parameters will include properties of personality, physicality, and ambient environmental conditions.

To be clear on the notation of Eq.~\ref{eq:ising} versus Eq.~\ref{eq:ODE}: each of the $N$ total entities $\bm{x}_i$ evolves in $A$ state variables $x_a$:
\begin{equation*}
  \bm{x}_i = [x_{i,1}, x_{i,2}, ..., x_{i,A}].
\end{equation*}
\noindent Then the total population $\bm{X}$ is: 
\begin{align*}
  \bm{X} = \sum_i{\bm{x}_i} = &[x_{1,1}, x_{1,2}, ..., x_{1,A}, x_{2,1}, x_{2,2}, ..., x_{2,A}, \\ 
  & ..., x_{N,1}, x_{N,2}, ..., x_{N,A}].
\end{align*}
\noindent We then rewrite $\bm{X}$ in B state variables $x_b$:
\begin{equation*}
  \bm{X} = [x_1, x_2, x_3, ... x_B],
\end{equation*}
\noindent where $B$ -- the dimension of Eq.~\ref{eq:ODE} -- equals the number of entities $N$ multiplied by the dimension $A$ of each entity, and we have taken $A$ to be invariant across entities.  For simplicity, hereafter we shall drop the subscripts on $\bm{R}$ and $\bm{S}$.

Now, in neurobiology, the $\bm{x}_i$ are neurons.  In that context, the self term $\bm{R}(\bm{x}_i)$ describes intrinsic electrophysiological properties of Neuron $i$~\cite{hodgkin1952quantitative}, and the interaction term $\bm{S}(\bm{x}_i,\bm{x}_j)$ describes the effect upon Neuron $i$ due to Neuron $j$ (for the specific form of $\bm{S}$ in that context, see Appendix B).  One state variable of great interest is a neuron's membrane voltage, a measure of cellular activity.  If an interaction term from Neuron $j$ to Neuron $i$ is sufficiently positive ("excitatory"), then Neuron $i$ will fire an electrical impulse in response to its interaction with Neuron $j$\footnote{For comedians, think of excitatory input as audience laughter.}.  On the other hand, sufficiently negative ("inhibitory") values of $\bm{S}(\bm{x}_i,\bm{x}_j)$ will suppress the membrane voltage dynamics of Neuron $i$.  In the extreme case, this suppression is permanent. 

Let us take the electrical activity of interest to be cardiovascular in nature.  Specifically, I seek to send the dynamics of state variable $x_{\text{Conan},\text{heart}}$ to zero, by setting interaction term $\bm{S}(\bm{x}_{Conan},\bm{x}_{me})$ to a sufficiently negative value.  And I shall seek to do that in a manner such that my state variables $x_{me,a}$ (a: 1,2,...,A) cannot be implicated uniquely in achieving that aim.

As with the measurements, there is ample opportunity in the model to create confusion.  Entities $\bm{x}_i$ will be defined by numerous physiological\footnote{For example, it could be important to know whether the person has hands.} and psychological properties, and so the dimension $B$ can be formidably high.  Second, the specific forms of $\bm{R}$ and $\bm{S}$ of Eq.~\ref{eq:ising} are nonlinear and poorly constrained by theory. 

For a taste of nonlinearity, imagine that two of the state variables that define me are $x_{me,neuroticism}$ and $x_{me,conscientiousness}$.  If both variables are dynamically active, then the self term $\bm{R}(\bm{x}_{me})$ that operates on them may indicate that, while I harbor multiple unsavory tendencies, I am sufficiently civilized to minimize the negative impact on those around me\footnote{Say, I turn my neuroticism inward and deride myself in the bathroom mirror multiple times a day.}:  
\begin{align} \label{eq:R1}
  \bm{R}&(\bm{x}_{\text{me,neuroticism}}=\text{HIGH},\bm{x}_{\text{me,conscientiousness}}=\text{HIGH})\\\nonumber &\rightarrow \text{"troubled but harmless to others besides self."}
\end{align}  

On the other hand, if the dynamics of $x_{me,neuroticism}$ remain highly active and those of $x_{me,conscientiousness}$ are low, then self term $\bm{R}(\bm{x}_{me})$ may render my disposition innately murdery toward others besides myself:
\begin{align} \label{eq:R2}
  \bm{R}&(\bm{x}_{\text{me,neuroticism}}=\text{HIGH},\bm{x}_{\text{me,conscientiousness}}=\text{LOW})\\\nonumber &\rightarrow \text{"murdery."}
\end{align}  
\noindent The relative weightings of those two state variables $\bm{x}_{\text{me,neuroticism}}$ and $\bm{x}_{\text{me,conscientiousness}}$ will be set by the parameters $\bm{p}$ that define the specific form of $\bm{R}$. The values of these parameters must be inferred from the available measurements -- and a realistic model may contain hundreds to thousands of parameters\footnote{Take this personality quiz~\cite{quiz} and you'll see what I mean.}.  

Still more gnarly than self term $\bm{R}$: interaction term $\bm{S}$ will consider my state variables \textit{together with} those of Conan O'Brien.  It's a beastly nonlinear problem, with plenty of places for a murderer to hide.

\section{Optimization} \label{sec:opt}

Criminal investigators will seek to infer the dynamical evolution of the full model population $\bm{X}$ back through the time of the murder -- a solution that will identify those state variables $x_{i,a}$ belonging to entity $\bm{x}_i$ who is most likely responsible for having sent the dynamics of Conan O'Brien's heart rate ($x_{\text{Conan},\text{heart}}$) to zero.  To thwart their efforts, we must first understand how they work.

Let us assume that this murder investigation will take the form of Bayesian inference~\cite{von2011bayesian}, a statistical method used to update the probability for a hypothesis as evidence accumulates.  The specific formulation used in this paper was invented for numerical weather prediction~\cite{kalnay2003atmospheric}; it is designed for the case in which available measurements are sparse, as is often the case in murder investigations.  

Inference can be formulated as an optimization, wherein solutions must "optimize" measurements with model: they must align with both any available measurements, and any constraints that can be placed on the model dynamics.  These requirements are imposed by means of extremizing a cost function $C$:
\begin{widetext}
\begin{equation}
\label{eq:cost}
  C = \underbrace{ \mathlarger{\sum}_{\text{t=u}} \mathlarger{\sum}_{l=1}^L ({\color{crimson} m}_l(t_u) - \bm{h}_{l,u}(x_l(t_u)))^2}_{\text{measurement error}} + \underbrace{\sum_{\text{t=v}}^{V-1}\sum_{b=1}^{B} \left(x_b(t_{v+1}) - F_b(\bm{X}(t_v),\bm{p}(t_v))\right)^2}_{\text{model error}},
\end{equation}
\end{widetext}
%\begin{equation*}
% A = \vertarrowbox{2b}{\text{stuff}}}
%\end{equation*}
\noindent where the "measurement" and "model error" terms penalize a solution's deviation from measurements and model dynamics, respectively.  In the measurement error term, the inner sum considers all measurements ${\color{crimson} m}_l$, of which there are $L$ (location and time, to name two).  Then $\bm{h}$ is the transfer function, or map, between the ${\color{crimson} m}_l$ and associated model state variables $x_l$.  The outer sum is on time, and enforces the measurements ${\color{crimson} m}_l$ at temporal locations $u$ at which they were obtained.  

The model-error term penalizes deviation from known dynamics.  The inner sum considers all $B$ state variables -- both measured and unmeasured.  The outer sum on time enforces the term at all discretized temporal locations $v$ of the model, a set that contains all measurement locations $u$.  Because both terms are extremized simultaneously, information in the measurements ${\color{crimson} m}_l$ can propagate to the model-error term to estimate the unknown parameters $\bm{p}$. 

The transfer function $\bm{h}$ is important (recall Fig.~\ref{fig1}): it translates the measurements to a murderer profile.  Take a simple case with $L=3$ and $B=3$, at one temporal location $u$:
\begin{equation}
\label{h}  
\left(\begin{array}{c}
     {\color{crimson} m}_{\text{location}} \\
  {{\color{crimson} m}_{\text{time}}} \\
  {{\color{crimson} m}_{\text{weapon}}} %  & 
\end{array} \right)  = \bm{h}_{l,u}
 \left(\begin{array}{c}
     x_{i,\text{neuroticism}} \\
  {x_{i,\text{conscientiousness}}} \\
  {x_{i,\text{impulsivity}}} % & 
\end{array} \right).
\end{equation}
\noindent Cracking the specific form of $\bm{h}$ will fall under the purview of the investigators.  For example, they might find themselves asking: "What does the measurement ${\color{crimson} m}_{weapon} = \text{"Emmy trophy"}$~\cite{emmy} indicate about the murderer's state variable $x_{i,\textit{impulsivity}}$?"  Although as the murderer, the determination of $\bm{h}$ is out of your hands, take heart: writing $\bm{h}$ will be even gnarlier than inferring the model terms $\bm{R}$ and $\bm{S}$.  This is because investigators will have to represent any \textit{un}measured quantities as a superposition of amplitudes for existing across all possible states\footnote{For example, if Conan O'Brien's body is missing, then the murderee is a superposition of amplitudes representing all possibly-murdered people, living or dead.} -- and most degree programs in forensics overlook quantum mechanics in their curricula.  Hence most investigators are ill-prepared to constrain $\bm{h}$.

Successful estimations of all unknown model parameters $\bm{p}$ will permit investigators to predict the evolution of Population $\bm{X}$ back to the temporal window of the murder.  Of particular interest are those parameters that send interaction terms $\bm{S}(\bm{x}_{Conan},\bm{x}_j)$ to values sufficiently negative to be fatal, thereby identifying entities $\bm{x}_j$ as murderers.  It is a $(B + p) \times(V + 1)$-dimensional solution\footnote{Regarding the choice of temporal step size: if the murder occurred quickly, one must incur computational expense in order to resolve the dynamics.  In addition, the farther into the past the murder occurred, the more steps will be required to rewind to it.  On the other hand, if the time of death ${\color{crimson} m}_{time}$ is known precisely, then investigators may save computational expense by using an adaptive step: large outside the window of interest and increasingly fine closer in.}.  For a short derivation of Eq.~\ref{eq:cost}, see Appendix A.

It is best practice to perform this procedure multiple times in parallel computations, where each initialization is an independent and random guess at the true solution.  A robust result is that in which all initializations converge to the same solution.

\section{Sabotaging the Optimization} \label{sec:sabotage}

The theory of optimization and control is dedicated to minimizing measurement noise and maximizing model resolution, to identify one "global minimum" of the cost function that is consistent with both measurements and model.  Let us now torpedo this aim.  

I shall present two possibilities: 1) creating degeneracy, and 2) rendering the measurements incompatible with a model in which I am the murderer.   We shall explore each in turn.

\subsection{Create degeneracy}
 
In this scenario, let us accept that there exists at least one solution to Conan O'Brien's murder in which I am the murderer. 
 I aim to create additional (degenerate) solutions that contain alternative murderers.  The more I can gin up, the lower the probability that the authorities will ultimately settle on me\footnote{A warning to lazy murderers, who may be thinking that the model is sufficiently nonlinear that multiple solutions (local minima) will crop up anyway on the cost function surface, and so you needn't do more: powerful computational techniques exist to deal with local minima (e.g. Ref.~\cite{van1989bayesian}).  I advise you to further complicate things as much as you can.}.  

Now, a caution: there exists no guarantee that the authorities will \textit{not} eventually settle on me.  Further, so long as they are unable to settle on anyone, the case remains unsolved -- leaving time for the emergence of additional measurements for which I am unprepared.  As we proceed with this avenue of sabotage, keep in mind the double-edged sword.

Formally, the aim is to permit as many interaction terms $\bm{S}(\bm{x}_{\text{Conan}},\bm{x}_{\text{j}})$ as I can to be as pathologically negative as my own $\bm{S}(\bm{x}_{\text{Conan}},\bm{x}_{\text{me}})$.  That is: those entities must be just as likely to send the dynamics of $x_{Conan,heart}$ to zero as I am.  Two general methods exist to achieve this: maximizing noise in measurements, and maximizing uncertainty on model parameter values. 

\subsubsection{Degeneracy via measurements}

Consider the linear equation "$Z = P Y$," where parameter $P$ has a known value of 1.  That is, the "parameter regime" is known precisely.  Say we measure $Y$ to be 4.  Then $Z = 1 \times 4 = 4$, a unique solution.  

\begin{figure*}[htb]  
  \includegraphics[width=\textwidth]{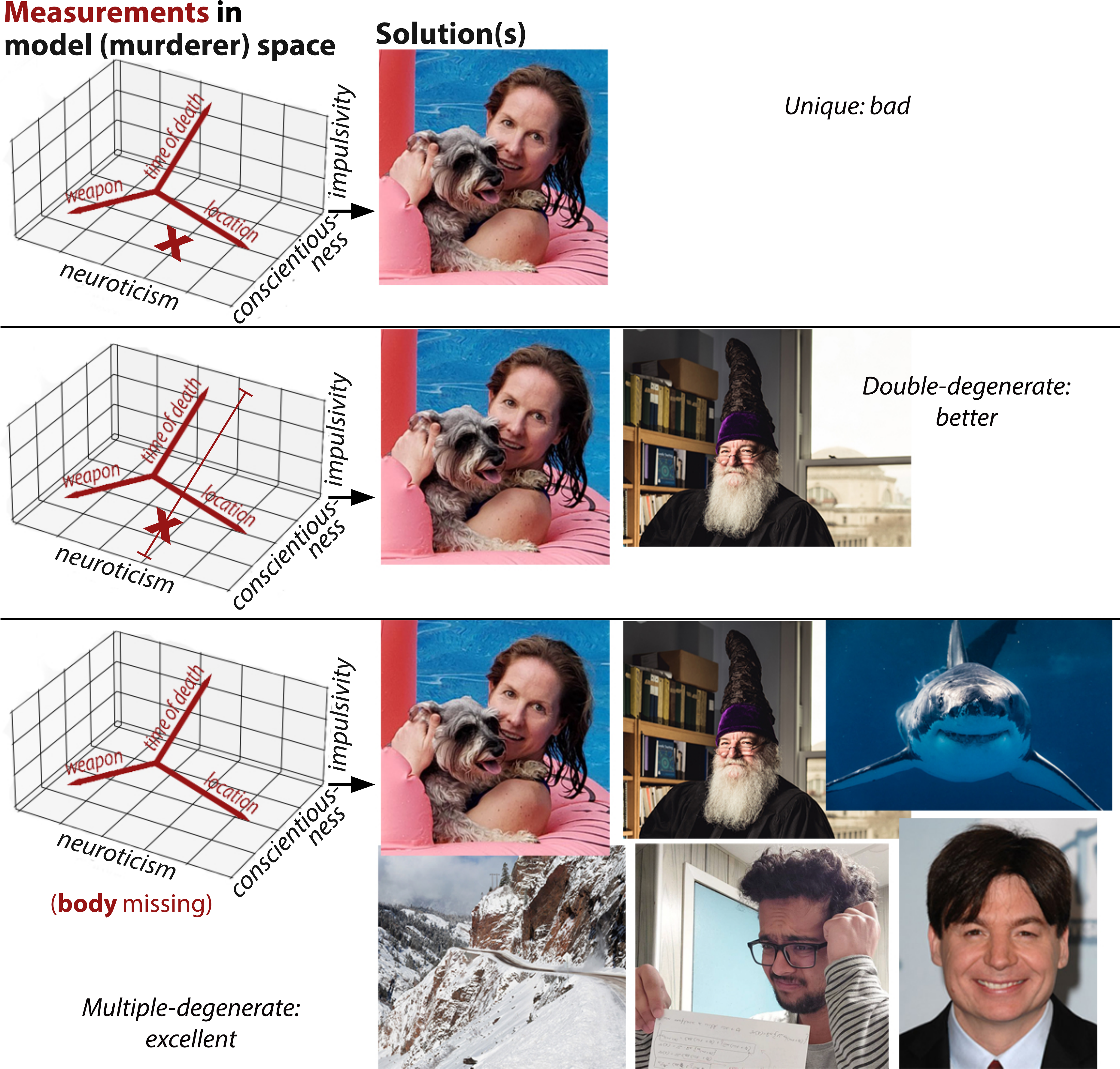}
  \caption{\textbf{Precision of measurements X versus number of solutions.}  \textit{Left}: Measurements X (crimson) in model space (black); \textit{right}: murderers permitted.  \textit{Top}: Measurements X are complete and with zero errors, permitting a unique solution wherein the murderer is me (but note strategic inclusion of cute puppy~\cite{gretel}.)  \textit{Middle}: Measurement of time-of-death contains large uncertainty, which permits a second murderer: esteemed  -- and framed -- professor of astrophysics at Columbia University, David Helfand~\cite{helfand}.  \textit{Bottom}: Measurement of body is missing, so that $\mathcal O(10^4)$ solutions are permitted.  Six murderers representative of the complete set are shown; clockwise from David Helfand: shark~\cite{shark}, comedian/actor Mike Myers~\cite{mikeMyers}, severely disgruntled Ph.D. student~\cite{ishaan}, icy road near cliff~\cite{icyRoad} (none of whom have cute puppies.)} \label{fig3}
\end{figure*}

Fig.~\ref{fig3} depicts an analogous murder scenario.  At left are the measurements (crimson) in model space (black), as per Fig.~\ref{fig1}.  In the top panel, the crimson "X" denotes a set of measurements that are complete and precisely known.  For example, the murder was caught on surveillance video.  At right are the associated solutions obtained by the optimization.  The impeccable real-time data readily cherry-picks a unique solution, wherein the murderer is $\bm{x}_{me}$\footnote{Note that I strategically depict myself in the company of a cute puppy~\cite{gretel}.  People with cute puppies are perceived as more trustworthy than those without cute puppies~\cite{golbeck2010trust}.}.  Not good.  

Now let's rewrite the equation to be nonlinear: "$Z = P \sqrt{Y}$," again within the precisely known parameter regime of $P = 1$, and with $Y = 4.$  Then $Z = +2$ \textit{or} $-2$: two degenerate solutions.  

An analogous murder tactic may be to introduce uncertainty into the time of death ${\color{crimson} m}_{\text{time}}$.  At left in the middle panel of Fig.~\ref{fig3}, note the large error bars on the "time" axis in measurement space.  As a result, a second solution emerges in which the murderer is $\bm{x}_{\text{David Helfand}}$.  David Helfand~\cite{helfand} is an esteemed professor of astrophysics at Columbia University, author of "A Survival Guide to the Misinformation Age"~\cite{helfand2016survival}, and my former undergraduate adviser. In introducing uncertainty into measurement ${\color{crimson} m}_{\text{time}}$\footnote{Adding high uncertainty into time of death has the added benefit of precluding the investigators' use of an adaptive step size, thereby forcing them to incur computational expense.  With some luck, you might even crash their simulation.}, I have successfully framed David Helfand for murder\footnote{For an example of how I might have done this, see Sec.~\ref{subsub:degenModel}.}.

Generally, the more nonlinear the problem, and/or the sparser and noisier the measurements, the more solutions are permitted.  The bottom panel of Fig.~\ref{fig3} depicts a representative handful of the myriad solutions that exist in the ideal scenario wherein I destroy the body\footnote{Destroying a body is hard.  For a terrific tutorial, watch \textit{The Sopranos}, Episode 9 of Season 4~\cite{sopranos}.}.  I remain among the solutions, but now I am camouflaged by $\mathcal O(10^4)$ other possibilities.  In addition to David Helfand, other framees include a shark~\cite{shark}, comedian/actor Mike Myers~\cite{mikeMyers}, a severely disgruntled Ph.D. student~\cite{ishaan}, and an icy road near a cliff~\cite{icyRoad}; together they are representative of the complete set\footnote{It has been shown by Agatha Christie that one can even cram multiple murderers into a single solution~\cite{christie2018murder}.  The sky's the limit.}.  For more examples on introducing degeneracy via measurements, see Table~\ref{table1}, top right.

\subsubsection{Degeneracy via model} \label{subsub:degenModel}

Let's reexamine the equation "$Z = P \sqrt{Y}$," this time supposing that the value of parameter $P$ is not known to be precisely 1.  Rather, $P$  is in the ballpark of 1 to 1,000,000,000: the parameter regime is poorly constrained.  Then even if measurement $Y$ is known precisely, the solution $Z$ can take on myriad values.  

In the murder scenario, the main parameters of interest are those that define interaction term $\bm{S}$ of Eq.~\ref{eq:ising}.  As noted, the more negative a suspect's interaction term with Conan O'Brien, the more suspect the suspect.

Parameter estimation is notoriously treacherous compared to state variable estimation alone~\cite{carrassi2011state}, chiefly because parameters do not obey a known dynamical law.  Thus there exists no straightforward way to correlate errors in parameter values with either the evolution of state variables or with measurements.  Further, the parameters may themselves be time-varying\footnote{But most investigators will take many environmental parameters to be constant throughout the murder.  For example, the identity of my mother.},\footnote{Not that this murder scheme has anything to do with my mother.}. 
% \footnote{\sout{That correlation will depend on the model sensitivity to the parameter in question, which, at first order, is proportional to the Jacobian of the model with respect to that parameter.}}

More good news: theories of personality and social dynamics are extremely diverse and varied.  Perspectives include dispositional, psychodynamic, humanistic, genetic, behavioral, social-learning, and cognitive~\cite{schultz2016theories}.  So not only does there exist disagreement on how to write the specific forms of terms $\bm{R}$ and $\bm{S}$ governing the state variables, but there exists no consensus on what those state variables should even be\footnote{For example, some question whether "neuroticism" and "impulsivity" (Fig.~\ref{fig1}) indeed define orthogonal directions in the model space~\cite{valero2014neuroticism}.}.  

Here are a few tactics I might employ to obscure constraints on model parameters.  Shortly before the murder, I suggest to David Helfand that he seems stressed and should go for a long solitary bike ride in Riverside Park.  Thus I am increasing the probability that he will lack an alibi for the time of the murder.  Mathematically, I am loosening constraints on parameters governing the interaction term $\bm{S}(\bm{x}_{\text{Conan}},\bm{x}_{\text{Helfand}})$.  Alternatively, I might manipulate the self term $\bm{R}(\bm{x}_{\text{Helfand}})$: shortly before the murder, I insult his cooking.  Thus I have primed him to enter a bad mood so that witnesses will describe him as "rather murdery" around the time of the murder. 

My best general advice is to study murder statistics and act accordingly.  For example, at least 20\% of murders are committed by a relative of the victim~\cite{durose2005family}.  So try framing one of Conan O'Brien's siblings.  Remember that investigators will gravitate to the most probable explanation\footnote{So it is fitting that ours is a Bayesian methodology.}.  For other model-based examples on creating degeneracy, see Table~\ref{table1} top left.

\subsection{Eliminate solutions in which the murderer is me}

An alternative to creating degeneracy is to minimize the probability that solutions exist in which my state variables $\bm{x}_{me,a}$ are responsible for sending the dynamics of $x_{\text{Conan,heart}}$ to zero.  If in addition I can frame somebody else uniquely, this technique has the potential to offer vastly more closure than the degeneracy strategy.  As with degeneracy, there are two general approaches here: via measurements and via model.

\subsubsection{Eliminate self via measurements}

Strategies here involve adding fake noise, or faking the measurements entirely -- along the lines of kidnapping a cute puppy.  For more examples, see Table~\ref{table1} bottom right.

\subsubsection{Eliminate self via model}

With this route, the aim is to send my interaction term with Conan O'Brien $\bm{S}(\bm{x}_{\text{Conan}},\bm{x}_{\text{me}})$ to as positive a value as possible.  This involves faking my parameter

\setlength{\tabcolsep}{5pt}
\begin{table*}[!htbp]
\small
\centering
\begin{tabular}{p{2cm}|p{3cm} p{3cm}|p{3cm} p{3cm}}
\toprule
%\begin{tabular}{c | c  c | c c} \toprule
 Aim & Means: \textit{Model} & & \textit{Measurements} & \\\midrule \hline
 & \textit{Example} & \textit{Intended confusion} & \textit{Example} & \textit{Intended confusion}\\ \cline{2-5}
 \textit{Create degenerate solutions} & Shortly before I do it, steal the cupcakes from the graduate student lounge and tell them Conan O'Brien took them (\textit{i.e. in interaction terms $\bm{S}$ of Eq.~\ref{eq:ising}, I set values of environmental parameters to describe a cupcake-less environment}.) & The students are more likely to behave antagonistically -- and thereby suspiciously -- toward Conan O'Brien shortly before his demise.  & Destroy  body\footnote{Difficult, as noted.  See Footnote 19.} (\textit{i.e. add infinite noise to the ${\color{crimson} m}_l$ that are body-related}.) & This renders the person simply missing -- an extremely weak measurement, which in most states will buy me seven years~\cite{LII}. \\\cline{2-5}
 & Shortly before I do it, send as many other colleagues as I can on long solitary bike rides\footnote{I must send them to different places so they don't run into each other.  And remember that David Helfand might already be in Riverside Park.} (\textit{loosen constraints on state variables representing physical activity}.) & This lowers the probability that they will have alibis for the time of my murder.  & Freeze or burn body (\textit{add noise to ${\color{crimson} m}_{time}$}.) & This obscures time of death.\\\midrule \hline\hline
 \textit{Eliminate me-as-murderer from solutions} & When someone mentions that Conan O'Brien has been murdered, act surprised and say, "I thought he already got murdered in self defense by Mike Myers~\cite{mikeMyers}, like, five years ago" (\textit{fake-set my interaction term $\bm{S}$ with Conan O'Brien to zero for the past five years}.) & I appear to lack motive. & Rejigger time of death to Year 2046 (\textit{fake ${\color{crimson} m}_{time}$}.) & No documentation yet exists that I have experienced the year 2046. \\\cline{2-5}
 & The morning after the murder, submit a completed NSF grant proposal (\textit{fake tight constraints on my self term $\bm{R}$}.) &  I appear to lack opportunity: clearly, I have not left my desk for the past six weeks. & Move body to Gstaad, Switzerland (\textit{fake ${\color{crimson} m}_{location}$}.) & I have no documented association with Gstaad, Switzerland.\\\cline{2-5}
 & Cut off own hands (\textit{send dynamics of $x_{me,hands}$ to zero; hope they will be assumed constant back through the murder}.) & I appear to lack means. & Replace body with body of Jane Lynch~\cite{janeLynch} (\textit{fake the ${\color{crimson} m}_l$ that define body's identity}.) & I had no documented beef with Jane Lynch.\\\bottomrule
\end{tabular}
\caption{\textbf{Computational strategies for getting away with murdering Conan O'Brien.}  \textit{Top}: By creating degeneracy; \textit{bottom}: by rendering all solutions incompatible with me-as-murderer.  Each category offers options to achieve the aim by tampering with the model (left) or  measurements (right).}
\label{table1}
\end{table*}

\setlength{\tabcolsep}{5pt}
\begin{table*}[!htbp]
\small
\centering
\begin{tabular}{p{4cm}|p{6cm}|p{6cm}} \toprule
%\begin{tabular}{c | c  c | c c} \toprule
 Situation & Tip & Mathematical implementation \\\midrule \hline
 You seek to murder a family member. & Family members are obvious suspects~\cite{durose2005family}, so do as much as you can here.  Act unremarkably pleasant, acquire a cute puppy (\`a la Fig.~\ref{fig3} top middle), and get rid of the body or at least make it look like an accident (Fig.~\ref{fig3} bottom middle).  And try reverse psychology: make comments like, "Hey, wouldn't I be a terrific murderer?  Ha ha ha!"\footnote{This is how self-deprecating humor can function as a bully shield: you pre-rob others of the glory that might otherwise come with making fun of you.}  & Fake model parameters governing self and interaction terms, and fake or erase as many measurements ${\color{crimson} m}_l$ as possible. \\\hline
 You've developed murder as a habit. &  The main risk with serial murder is that you're inadvertently leaving patterned measurements that will implicate you.  So render measurements sparse and inconsistent. & For example, don't have ${\color{crimson} m}_{\text{weapon}}$ consistently be "Emmy trophy."\\\hline
 You are considering murder for hire. & Determine whether it's worth it.  How much are they paying you?  And feel them out: does it have to be murder, or could you instead move the person alive to Gstaad, Switzerland? & Before you do it, run the optimization a few times using \textit{simulated} measurements.  Note the degeneracy present in solutions, to get a sense for how likely it is you'll pull it off. \\\hline 
 You completed the murder but dropped your wallet at the scene. &  Go back and get it. & Forget optimization, just go.  Hurry.\\\bottomrule
\end{tabular}
\caption{\textbf{Tips for prospective murderers on historically thorny situations.}}
\label{table2}
\end{table*}

\noindent 
regime.  It's tricky; it requires good acting.  First I might study literature on criminal profiles~\cite{douglas1986criminal} and be sure I do not fit them.  And within the Big Five theory of personalities~\cite{de2000big}, I'd practice scoring low on "neuroticism" and high on "agreeableness"\footnote{Ted Bundy pulled this off beautifully, for a while~\cite{michaud2000ted}.}.  Then, if the murderee is an experimentalist, I might stroll around professing my reverence for experimentalists.  I'd certainly hide my consternation regarding his cat noises.  

If rendering $\bm{S}$ positive winds up being too much of a stretch, it would even help to send it to zero, i.e. "prune" myself out of the model\footnote{Investigators will seek to prune the model anyway, to simplify things.  For example, if one possible murderer is an icy road near a cliff (Fig.~\ref{fig3}, bottom middle), they will probably take the road's state variable $x_{\text{road,conscientiousness}}$ to lack dynamics.}.  The main concern is that $\bm{S}$ not appear to be negative.  For more examples, see Table~\ref{table1} bottom left.
% I might feign no care in the world regarding tenure {\color{purple} it's been a while since i brought up tenure}, declaring that I'm certain to get it based on teaching evaluations alone.  

Note that the examples of Table~\ref{table1} are not mutually exclusive.  I could try manipulating measurements and model simultaneously; the more confusion the better.  In addition, Table~\ref{table2} offers computational tips for prospective murderers who are facing historically thorny situations.

\subsection{Crash the entire computation}

One final approach is to make the optimization procedure fail to converge on any solution whatsoever.  As with creating degeneracy, this tactic will have the undesirable effect of leaving the case dangling open, and it may be even trickier.  So I advise it only for those seeking a real challenge -- perhaps seasoned serial murderers who have grown bored.  

For you, Dear Readers, one idea: create highly precise (i.e. low-noise) measurements that are contradictory.  Fudge ${\color{crimson} m}_{\text{weapon}}$ and ${\color{crimson} m}_{\text{location}}$ to be "insufficient sunscreen" and "Seattle, Washington," respectively.  If each is taken to be known precisely, no wiggle room exists for the computations to reconcile them.  Remember, proceed with caution: the longer this contradiction thwarts the investigation, the more time you offer investigators to sniff the true location back to Los Angeles, California. 

\section{Result} \label{sec:result} 

See Footnote \footnote{Alas, it is an unconscionable feature of the scientific literature that null results go unreported.}

\section{Discussion} \label{sec:disc}

I have aimed to share optimization-based strategies for dodging murder conviction, and have done so via a series of opaque\footnote{Opaqueness aside for a moment: will somebody please go check that David Helfand's all right down in Riverside Park?  Tell him he can come back now.} examples.  Seasoned readers of the scientific literature may deride my lack of rigor: I have provided no statistics linking the magnitude of a particular measurement error to the number of resulting degenerate solutions, nor have I offered a specific form for Eq.~\ref{eq:ising}.  Forgive me; I am seeking to dodge arrest.  Further, I have chiefly written this manuscript for the subset of scientific researchers who are also prospective murderers.  You, my brethren, no doubt will understand my caginess.

There lurks something still cagier within the nonlinear model of Eq.~\ref{eq:ising}, which we have not yet examined: feedback.  While my interaction term with Conan O'Brien $\bm{S}(\bm{x}_{\text{Conan}},\bm{x}_{\text{me}})$ describes the manner in which I interact with Conan O'Brien, simultaneously Conan O'Brien's term $\bm{S}(\bm{x}_{\text{me}},\bm{x}_{\text{Conan}})$ describes the manner in which he interacts with me.  The two are evaluated simultaneously, yet they feed back upon each other.  And feedback is a likely recipe for chaos.

The term "chaos" describes the apparently random behavior of a dynamical system that emerges in some parameter regimes, due to the system's extreme sensitivity to small changes in its initial conditions.  Imagine you host a talk show and it's going pretty well, but then someone shifts your initial conditions by 30 minutes.  In this regime you might not know what dynamics will emerge.  Chaotic systems are inherently unpredictable. 
% Imagine that your blind date shows up with a puppy.  You might find that cute.  Now apply a minor perturbation so that the puppy is headless.  In this regime, you might not know how you'd react

Now recall from Eq.~\ref{eq:ising} that an extremely negative value of an interaction term $\bm{S}$ equates to murder, while an extremely positive value may be, say, love.  If chaos lurks within the form of $\bm{S}$, love may turn to murderous rage essentially instantaneously~\cite{graham2022explanatory}.  Do you not at times want to hug your mother, and at other times feel relieved that there isn't a meat cleaver handy\footnote{Not that this manuscript has anything to do with my mother.  I already said that.}?

We are all prospective murderers, Dear Reader.  Those dynamics are born within us: strong sensitivity to the current state is a tool for successfully navigating an ever-changing environment.  For our dynamics to go pathologically awry merely requires the appropriate tweak to our parameter regime.  In some universe -- some \textit{physically realistic} universe, Dear Reader -- \textit{you} murdered Conan O'Brien.  We all did.\\

\begin{centering}
\hspace{4.2cm}\huge{$\sim$}
\end{centering}
%\subsection{Using off-diagonal terms}  {\color{purple} I ALREADY PUT AN ABRIDGED VERSION OF THIS IN EARLIER.  Search "quantum mechanics" to find it.}
%You may have noted that the transfer function $h$ may be a matrix of dimension as large as X.  Permitting off-diagonal terms in matrix governing meas to model: okay -- while the measurements exist in definite eigenstates, if the murderer is never caught -- and in the case where you successfully blow up the procedure, then there is no collapse to a model eigenstate.  OR: Meaning of lack of measurements: Essentially, if you have no body, then the murderee exists as a superposition of all possibly-murdered people.  This significantly broadens the pool of likely murderers.  PUT THIS INTO MAIN TEXT?

\section{Acknowledgements}

E.A. thanks the National Science Foundation for Grant 2139004, and comedian/writer/humanBeing Conan O'Brien for inspiration.

\section*{Appendix A}

The cost function of Eq.~\ref{eq:cost} can be derived from the physical principle of least action.  Attributed to Pierre Louis Maupertuis~\cite{deessai} (1744), Leonhard Euler~\cite{euler1744methodus} (1744), and Gottfried Leibniz~\cite{kabitz1913gotha} (early 1700's), this is a variational principle from which one can derive the Newtonian, Lagrangian, and Hamiltonian equations of motion of a dynamical system.  It is both a profound and useful statement: that among an infinite number of possible paths that a system can take in a state space, the path that the system takes is that which minimizes one scalar quantity: the action.  
% ,gerhardt1898ueber - for Leibniz

The principle of least action underlies path integral approaches to statistical data assimilation (SDA)~\cite{abarbanel2013predicting,restrepo2008path,kalnay2003atmospheric}.  SDA is the specific procedure used in this manuscript; I kept the name out of the main text lest it send the reader fleeing.  The methodology can ask: \textit{Which measurements must be made in order to estimate unknown model components, for predictive purposes?  Further, what information about the system is contained within a certain set of measurements?}  A path-integral-based approach to SDA distills the essence of inference to a governing principle across disciplines, and in this way, disjoint disciplines in fact possess commonality and can inform one another.  It is in this way that we can take wisdom gleaned from SDA applications in geophysics, neurobiology, and astrophysics (see references in Sec.~\ref{sec:intro}) and seamlessly apply it to murder.

The path integral is an integral representation of the master equation for the stochastic process  represented by Eq.~\ref{eq:ODE}.  We seek the probability of obtaining a path $\bm{X}$ in the model's state space given observations $\bm{Y}$: 
\begin{align*}
  P(\bm{X}|\bm{Y}) = e^{-A_0(\bm{X},\bm{Y})},
\end{align*}
\noindent
or: \textit{the path $\bm{X}$ for which the probability - given $\bm{Y}$ - is greatest is the path that minimizes the quantity $A_0$}, which we call our action.  A formulation for $A_0$ will permit us to obtain the expectation value of any function $G(\bm{X})$ on a path $\bm{X}$: 
\begin{align*}
  G(\bm{X}) = \langle G(\bm{X}) \rangle = \frac{\int d\bm{X} G(\bm{X}) e^{-A_0(\bm{X},\bm{Y})}}{\int d\bm{X} e^{-A_0(\bm{X},\bm{Y})}}, 
\end{align*}
\noindent
where in our case the quantity of interest is the path itself.  The action is written in two terms:
\begin{align}
  A_0(\bm{X},\bm{Y}) = -\mathlarger{\sum} \log[P(\bm{x}(n+1)|\bm{x}(n))] - \\\nonumber \mathlarger{\sum} \text{CMI}(\bm{x}(n),\bm{y}(n)|\bm{Y}(n-1)).
\label{eq:action}
\end{align}
\noindent The first term describes Markov-chain transition probabilities~\cite{markov1906extension} of observing state $\bm{x}$ at time $(n+1)$ given state $\bm{x}$ at time $(n)$.  The second term is the conditional mutual information~\cite{fano1961transmission} (CMI), which asks: \textit{How
much information, in bits, is learned about event $\bm{x}$(n)  upon observing event $\bm{y}$(n), conditioned on having previously observed event(s) $\bm{Y}$(n - 1)?}  Simplifications then reduce the Markov-chain and CMI terms, respectively, to the "model" and "measurement" error terms of Eq.~\ref{eq:cost}~\cite{abarbanel2013predicting}.  

Optimization is performed on Eq.~\ref{eq:cost} by requiring that small variations to the action vanish under small perturbations~\cite{oden2012variational}, thereby enforcing the Euler-Lagrange equations of motion upon any path.  This requirement ensures that all murderers, regardless of technique, obey the laws of physics.  I know that can be a nasty pill to swallow for my target audience: murderers are not fond of laws.  Well, what can I say -- this is the universe you picked\footnote{More accurately, this is the universe your mother picked for you.  So take it up with her.}. 
\hspace{0.5cm}

\section*{Appendix B: The interaction term in neuroscience}

For the interested reader, I offer an example from neuroscience of a specific form of the interaction term $\bm{S}(\bm{x}_i,\bm{x}_j)$ of Eq.~\ref{eq:ising}.
When considering two biological neurons $\bm{x}_i$ and $\bm{x}_j$, the term represents the electric current transmitted between the cells.  Specifically, $\bm{S}$ is the synaptic input current $I_{syn,ij}$ to Neuron $i$ from Neuron $j$:
\begin{equation} \label{eq:neuro}
  I_{syn,ij} = g_{ij}(t) w_{ij}(t)(E_{syn,ij} - V_{i}(t)).
%  \diff{s_{ij}}{t} &= T(t)[1 - s_{ij}(t)] - \beta s_{ij}(t)\\   
%  T(t) &= \frac{T_{max}}{1 + \exp(-(V_{pre}(t) - V_{P})/K_{P})} 
\end{equation}
\noindent
The function $g_{ij}(t)$ is the coupling strength between the neurons, and depends on neuromodulatory processes.  The function $w_{ij}(t)$ is the "gating variable": the fraction of pores in the membrane of Neuron $i$ that are open at any time to permit the current from Neuron $j$ to enter; for a specific form of $w$, see Ref.~\cite{destexhe1994synthesis}.  The parameter $E_{syn,ij}$ is the electric potential at which the net flow of charge through Neuron $i$'s membrane is zero, and $V_{i}(t)$ is the instantaneous membrane potential of Neuron $i$.  The intended takeaway: Eq.~\ref{eq:neuro} is gnarly.

For human interactions, the specific form of $\bm{S}$ is orders of magnitude more gnarly, and is beyond the scope of this paper.

%\bibliographystyle{acm}
%\nocite{*}
\nocite{TitlesOn}
\bibliography{bib_AF2023}
%\bibliography{bib_AF2023,bib_physics,refs_sciCom,refs_broaderImpacts,refs_armstrong,refs_extraLinks,bib_neutrinos,bib_COVIDPaper,refs_astroObs,refs_neuro,bib_people,bib_dataAssimilation,bib_collisionsPaper,bib_solar}

%\bibliographystyle{unsrt}

\end{document}